 \pdfoutput=1
 \documentclass[final,5p,times,twocolumn]{elsarticle}
 \pdfoutput=1
 \newcommand{\pf}[2]{\frac{\partial ,1}{\partial ,2}}
\newcommand{\pfs}[2]{\frac{\partial^2 ,1}{\partial ,2 ^2}}
\newcommand{\pft}[2]{\frac{\partial^3 ,1}{\partial ,2 ^3}}
\newcommand{\pderiv}[2]{\frac{\partial ,1}{\partial ,2}}

\usepackage{multirow}
\usepackage{tabu}
\usepackage{CJKutf8}
\usepackage[T1]{fontenc}
\usepackage[utf8]{inputenc}
\usepackage{graphics, amsmath}
\usepackage{graphicx,color}
\usepackage{epsfig}
\usepackage{amssymb}
\usepackage{amsthm}
\usepackage{setspace}
\usepackage{hyperref}

\usepackage{underscore}
\usepackage[dvipsnames]{xcolor}
\usepackage{multirow, makecell}
\usepackage{colortbl}
\usepackage{dashrule}
\usepackage{ehhline}

\usepackage{array}
\usepackage{booktabs}
\usepackage{caption}
\makeatletter
\g@addto@macro{\endtabular}{\gdef\rowfonttype{}}
\makeatother
\newcommand{\rowfonttype}{}
\newcommand{\rowfont}[1]{
	\noalign{\gdef\rowfonttype{#1}}}%
\newcolumntype{L}{>{\rowfonttype\strut}l}

\usepackage{array, makecell}
\usepackage{boldline}

\usepackage{colortbl}
\usepackage{hhline}

\DeclareUnicodeCharacter{202F}{\,}
\DeclareUnicodeCharacter{2212}{-}

\begin{document}

	\begin{frontmatter}
		
		\title{An investigation into the reliability of newly proposed MoSi$_2$N$_4$/WSi$_2$N$_4$ field-effect transistors: A monte carlo study}
		
		\author[1,2]{Zahra Shomali\corref{cor1}}
		\address[1]{Department of Physics, Faculty of Basic Sciences, Tarbiat Modares University, Tehran 14115-175, Iran}
		\address[2]{School of Physics, Institute for Research in Fundamental Sciences (IPM), Tehran 19395-5531, Iran}
		\cortext[cor1]{Corresponding author, Tel.: +98(21) 82883147.}
		\ead{shomali@modares.ac.ir}

		\begin{abstract}
			
			Recently, the two dimensional complex MA$_2$Z$_4$ structures have been suggested as suitable replacements for silicon channels in field-effect transistors (FETs). Specifically, two materials of MoSi$_2$N$_4$ and WSi$_2$N$_4$ due to their very desirable electrical and thermal properties are noticed. On the other hand, the reliability of transistors, which is determined by the maximum temperature they obtain during the performance, specifies the usefulness of the newly proposed channels for thermal management solution. In this work, the FETs, including MoSi$_2$N$_4$ and WSi$_2$N$_4$ channels, are investigated using Monte Carlo simulation of the phonon Boltzmann equation. In particular, the phonon analysis has been carried out to investigate the peak temperature rise. Our calculations confirm that MoSi$_2$N$_4$ and WSi$_2$N$_4$ present lower maximum temperature than the previously suggested candidate, the blue phosphorene (BP) which itself reaches a shallow temperature. Concretely, the phonon exploration shows that the competition between the dominant heat carrier velocity, and its related frequency settles the maximum temperature value. The material WSi$_2$N$_4$ with much more phonons in TA mode, with almost high velocity and relatively low-frequency, shows adequate thermal condition, and its peak temperature is very low, say 110 K, less than that of BP. The material MoSi$_2$N$_4$ attains the maximum temperature of only 10 K less than BP peak temperature. This behavior attributes to the dominant LA phonons which are fast but also have high frequency and consequently make the temperature get larger than that of the WSi$_2$N$_4$. In summary, WSi$_2$N$_4$, with very low peak temperature, and in the next step MoSi$_2$N$_4$, both with beneficial electrical/thermal properties, are suggested as very suitable candidates for producing more reliable FETs, fulfilling the thermal management.
			
		\end{abstract}
		
		
	\end{frontmatter}
	

	\section{Introduction}
	 \pdfoutput=1
	In 1954, the first commercial silicon transistor aiming for operating at high temperatures and lower production cost, was launched. Since then, the size of these types of transistors has been reduced such that the number in a dense integrated circuit has been doubled every two years. Currently, the number of transistors on a single chip has reached a vast number of one hundred billion \cite{YCHuang2017}. On the other hand, the size reduction of three-dimensional (3D) transistors results in less heat dissipation rate due to the increase in static power and the leakage current. As a solution, using two-dimensional (2D) semiconductors with almost zero leakage current has been suggested \cite{Ghazanfarian2015,Shomali2018,Acar2022}. In recent years, the development of field-effect transistors with low-dimensional material-based channels has drawn attention. 
	
	Aside from proper electrical condition, the thermal management of the FETs is also a crucial dispute.As usually, more than 90$\%$ of the initial energy is lost as heat, the most critical common thermal issue of the FETs is handling the self-heating procedure \cite{Sulima2007}. Hence, there have always been attempts to propose mechanisms and materials which reduce dissipation. Hundreds of millions of transistors assembled on a chip always have power problems. This causes the temperature of the chip to rise abnormally, which prevents the device from proper functionality. Heat generation is caused by many processes, such as Joule heating, current crowding, and thermoelectric. The Joule heating is reported to be the dominant self-heating for low-dimensional systems \cite{Grosse2011}. Moreover, the cooling technology has not caught up the transistor miniaturization, and many electronic devices have become impractical because of the formation of the millimeter-scale hotspots on the chip \cite{Balachandra1991,EPop2006,Pop2006,EPop2010,Ghazanfarian2009,Shomali2012,Samian2013,Samian2014,Moghadam2014,Shomali20152,Shomali2022,Valasa2022}. 
	
	Just as the chip-level hotspots are problematic \cite{Mahajan2002,EPop2005}, the thermal management issues at nanometer-length scales within individual transistors also take place. Considering that the newest 2D materials channels have lower thermal conductivities than bulk silicon, the new two-dimensional design of the FETs makes the heat spreading tough. In nano-electronics technology, the overall reliability is determined by the temperature of the hottest zone on the die. Consequently, the material selections and the heat spreader design \cite{Moore2014}, as a thermal management solution for controlling the temperature of these spots known as hotspots, are suggested. In addition to the attempts to find the better methods for heat removal, finding the low-dimensional silicon replacement candidates, with the lowest peak temperature rise, is the most proper and perfect choice for the nano-electronics industry. Rightfully so, the transistor with a lower maximum temperature can be easier kept under the threshold temperature \cite{Shomali2016,Shomali2018}. Accordingly, it is essential to investigate the thermal behavior of the electrically-efficient FETs, to ensure that they also cover energy efficiency for better cooling and operation.
	
	The FETs, including two-dimensional material channels like graphene, germanene, silicene, phosphorene allotropes, and Molybdenum disulfide (MoS$_2$), have been studied from electrical and thermal points of view. Each proposed candidate has advantages and disadvantages. For example, while graphene presents the lowest maximum temperature in response to the transistor self-heating, the absence of a band gap makes it an improper candidate. Aside graphene, two-dimensional materials with an acceptable band gap, resulting in a suitable on/off ratio, such as silicene, phosphorene allotropes, and Molybdenum disulfide (MoS$_2$) have been designed \cite{Chhowalla2016}. Apart from the appropriate band gap and on/off ratio, the mobility of the MoS$_2$, similar to the other transition metal dichalcogenides, is deficient. The mobility is increased using a high-k dielectric gate on top of it \cite{Shomali20152}, which makes the costs for using MoS$_2$ as channel high, and, consequently, introduces MoS$_2$ as an inappropriate candidate for newly developed field-effect transistor.
	
	Furthermore, silicene, and its sister material, germanene, as silicon replacement solutions have been widely studied. Silicene, which is the silicon-based analogue of graphene, has a band gap of 0.21 eV, and, accordingly an on/off ratio of 10. Also, the announced mobility of electrons and holes is about 1000 cm$^2$/V.S, which is low compared to that of graphene and blue phosphorene \cite{Lay2015,Tao2015} and more prominent than that of a single-layer MoS$_2$. On the other hand, germanene has a thin band gap which can be widened by applying an external electric field or by shrinking the material to nanoribbon. The advantage of germanene over silicene is having much higher mobility. Nevertheless, the experimental challenges in nanoribbons fabrication, make it not a perfect candidate for silicon replacement. Silicene circumstances with respect to germanene are more suitable as it is more compatible with the current silicon-based Microtechnology.
	
	Among the low-dimensional materials, phosphorene allotropes are obtained to be more suitable candidates. This is due to a high intrinsic bandgap (unlike graphene and germanene) and a high carrier mobility (in contradistinction to most transition-metal dichalcogenides). Above all desirable conditions, not all the phosphorene allotropes lead to efficacious thermal behavior. The anisotropic thermal properties of black phosphorene reduce the device's reliability and performance as the low thermal conductivity along the armchair direction causes the localized Joule heating. Out of all types of phosphorene allotropes, suggested as the channel, the blue phosphorene (BP), is reported as the favorable one. This is true while using BP as the channel is also thermally very efficient due to its isotropic thermal performance. The transistors, including BP channels, heat less and get lower temperature rise concerning self-heating effects \cite{Shomali2018}. In spite of all advantageous thermal and electrical properties, monolayer phosphorene has the disadvantage of instability under ambient conditions \cite{Jain2019}. Hence, the investigation to come across a stable thermally efficient monolayer 2D material with an admissible band gap and high carrier mobility, is still an open issue. 
	
	Lately, two new two-dimensional (2D) layered semiconducting materials, MoSi$_2$N$_4$ and WSi$_2$N$_4$, were experimentally synthesized. Further, a large family of such two 2D materials, called as MA$_2$Z$_4$, are also theoretically anticipated \cite{1intro,2intro,Yin2022,Shen2022}. These materials present properties that seem to cover the disadvantages of the previously proposed low-dimensional channels. Interestingly, MoSi$_2$N$_4$, and WSi$_2$N$_4$ have excellent ambient stability, moderate band gap, considerable carrier mobility and propitious thermal conductivity \cite{Hong2020,JYu2021}. In more detail, the calculated band gaps of MoSi$_2$N$_4$ and WSi$_2$N$_4$ are 1.73 eV and 2.06 eV, which are comparable to that of MoS$_2$'s \cite{QWang2020,Touski2021}. Furthermore, it is obtained that both monolayer MoSi$_2$N$_4$ and WSi$_2$N$_4$ have moderate electron mobility of around 100 cm$^2$ V$^{−1}$s$^{−1}$, which is much higher than 1-10 cm$^2$ V$^{−1}$s$^{−1}$ reported mobility for the single-layer MoS$_2$ transistors \cite{CLi2022,JHKim2016}. Moreover, although the complex crystal structure of Mo and W have heavy atomic mass, their thermal conductivity over a temperature range of 300 to 800 been found to be large. More particularly, at 300 K, the lattice thermal conductivity of MoSi$_2$N$_4$ and WSi$_2$N$_4$ is , respectively, 224 Wm$^{-1}$K$^{-1}$ and 219 Wm$^{-1}$K$^{-1}$.
	
	All these properties make MoSi$_2$N$_4$ and WSi$_2$N$_4$ single-layer materials, as electrically suitable candidates, for replacing silicon in FET channels. However, the perfectness of MA$_2$Z$_4$ for channel usage will be proved when they comprise relatively low maximum temperature during transistor functionality. In the present work, in search for figuring out the thermal efficiency of the FET, including single-layer MoSi$_2$N$_4$ and WSi$_2$N$_4$ channels, the peak temperature and the thermal heating/cooling of such field-effect transistors are investigated. The study is performed using non-equilibrium Monte Carlo simulation of the phonon Boltzmann equation. As the overall reliability and thermal efficiency of a FET are determined by the temperature of the hottest region of the transistor, applying the homogeneous heat flux of the heating generation zone, the maximum temperature for MoSi$_2$N$_4$ and WSi$_2$N$_4$ is calculated. Then, the thermal situation of single-layer MoSi$_2$N$_4$ and WSi$_2$N$_4$ FETs in comparison to BP single-layer low-dimensional FETs is investigated through phonon analysis. This paper is organized as follows. Section \ref{sec.2}, introduces the geometry and the boundary conditions. In Sections \ref{sec.3} and \ref{Numerical method}, the mathematical modeling and the numerical method are described. Sections \ref{Results} and \ref{conclusion}, respectively, present the results and conclusions.
	
	\section{Geometry and Boundary conditions}
	\label{sec.2}
	Heat transfer in two types of hexagonal non-centrosymmetric monolayer channels called MoSi$_2$N$_4$ and WSi$_2$N$_4$, are simulated. Due to the nanoscale effects, non-Fourier methods should be utilized \cite{Shomali2021}. Here, the calculations are performed using the Monte-Carlo simulation of the phonon Boltzmann equation. The studied area has width and length of L$_x$=200 nm and L$_y$=200 nm, with the thickness of 0.679 and 0.702 for MoSi$_2$N$_4$ and WSi$_2$N$_4$. At the center of the studied layers, the homogeneous heat generation volume of 30×200×0.53 nm is applied at the center of the channel. The depth d=0.53 nm, the thickness of BP, is considered to make the heat generation zone condition the same for all the materials. The heat source produces heat flux equal to Q=1$\times$10$^{19}$ W/m$^{3}$. All boundaries, except the bottom boundary, are adiabatic with temperature equal to the ambient temperature. The system also exchanges energy with the environment from the bottom boundary \cite{Shomali2017,2Shomali2017}. The initial temperature of all parts of the nano-devices is 299 K. Most of the generated heat inside the transistor spreads to the surroundings and toward the bottom. Further, the low-dimensional channels are contemplated as flat-edge (rectangularly shaped) with specular reflections in our simulation.
	
	\begin{figure}
		\label{Geometry}
		\hspace{-0.9cm}
		\centering
		\includegraphics[width=\columnwidth]{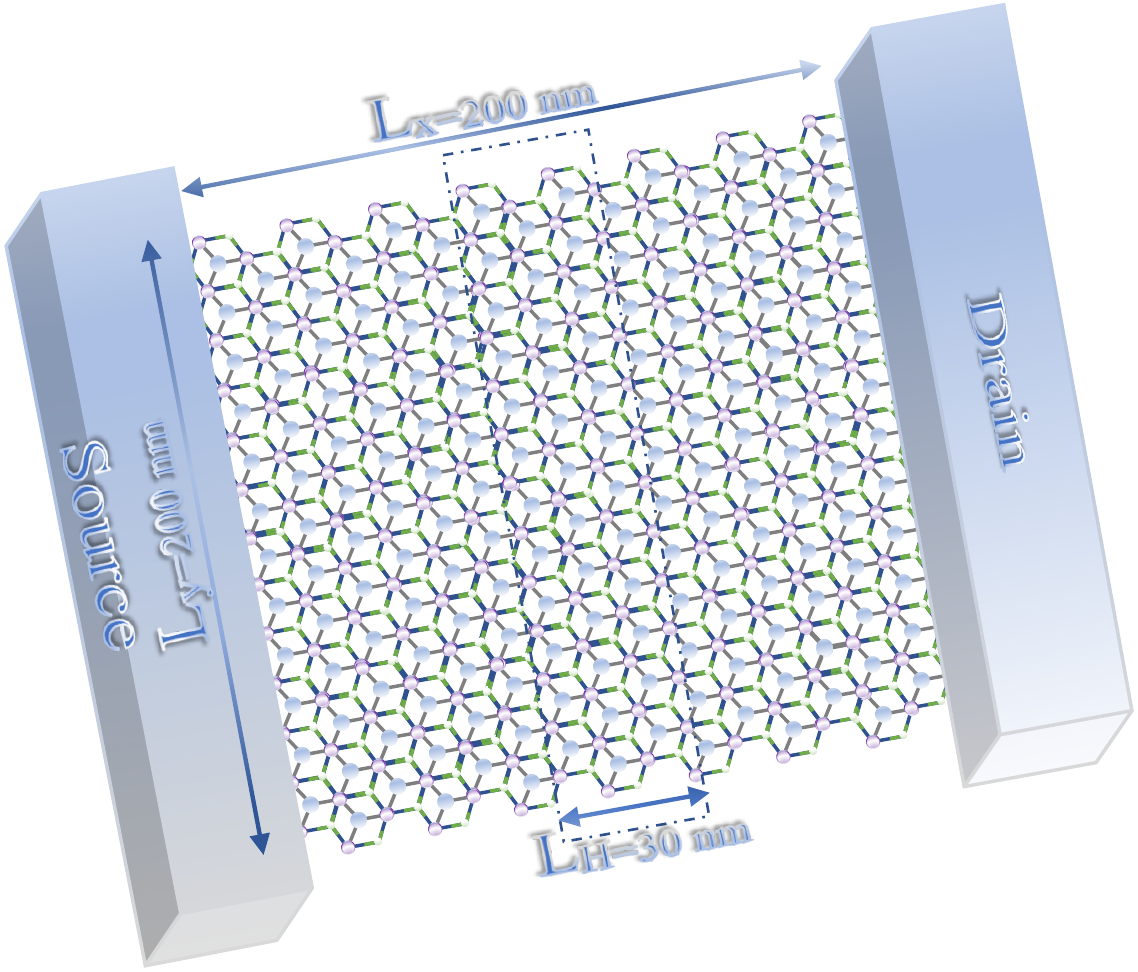}
		\caption{A typical MOSFET with a channel consisting of MoSi$_2$N$_4$ or WSi$_2$N$_4$. All boundaries are adiabatic except the bottom boundary. The parameters L$_x$, L$_y$ and L$_H$, respectively, present the length along the x, and y direction, and the width of the heat generation zone.}
	\end{figure}
	
	\section{Mathematical Modeling}
	\label{sec.3}
	
	Heat transport in materials is treated via the behavior of the phonons. Here, the phonon Boltzmann transport equation is applied to govern the phonons \cite{Shomali2017,Shomali2019,Sattler2022}. The equation is solved for determination of the phonon distribution function, f$_b$$(\mathbf{r},\mathbf{q},t)$, which depends on location, $\mathbf{r}$, wave vector, $\mathbf{q}$ and time t for each phonon branch, b. The phonon Boltzmann transport equation (PBTE) is written as 
	
	\begin{equation}
		\frac{\partial \textrm{f}_{b}(\mathbf{r},\mathbf{q},t)}{\partial t}+\textrm{v}_{b,\mathbf{q}} \ . \ \nabla_{r}\textrm{f}_{b}(\mathbf{r},\mathbf{q},t)=\frac{\partial \textrm{f}_b (\mathbf{r},\mathbf{q},t)}{\partial t}\mid_{scat},
	\end{equation}
	
	where v$_{b,\mathbf{q}}$, is the group velocity of phonons and depends on $\omega$$_{b,\mathbf{q}}$, the angular frequency of branch $b$ with the wave vector of $\mathbf{q}$. The parameter $\omega$$_{b,\mathbf{q}}$ relies on both the value and direction of the wave vector. Nevertheless here, the isotropic limit, in which the frequency is only value dependent, is investigated. In other words, large-area sheets of MoSi$_2$N$_4$, WSi$_2$N$_4$, and previously suggested blue phosphorene \cite{Shomali2018} low dimensional materials with almost isotropic phonon dispersion and consequently isotropic thermal conductivity are simulated \cite{THLiu2015,Zhang2017}. In this study, the stochastic phonon Monte Carlo (PMC) method is utilized to solve the PBTE. The most prominent inputs in PMC simulation are the phonon dispersion curves. These curves, reflecting the symmetry of the underlying lattice, present the relation between frequency $\omega$ and the wave vector $\vec{q}$ in the first Brillouin Zone (1BZ). From dispersion, the characteristics like group velocity and density of states are calculated. Then, the obtained results are used in the solution of BTE. The MoSi$_2$N$_4$, WSi$_2$N$_4$, and blue phosphorene, like other low-dimensional materials, has six phonon branches. The contribution of optical phonons at temperatures under 600 K is meager. Hence, only acoustic branches are taken into account in our simulation \cite{Mei2014}. Moreover, the quadratic plot is fitted to the full dispersion curves of all simulated low-dimensional materials \cite{Ge2016}. So, the relation $\omega$$_b$=c$_{b}$k$^{2}$+v$_{b}$k, is found for each low-dimensional material, and the calculated coefficients are shown in Table \ref{Tab1-tab1}.
	
	The simulation domain is a large area with the width, length, and thickness of W, L, and D. During the Monte Carlo simulation, phonons are considered as statistical samples, which are assigned to six individual stochastic spaces: three wave vectors and three position vectors. The sampled phonons initially undergo the drift (ballistic motion) and then go through to the scattering processes. The MC simulation steps for the low-dimensional material channel sheets are explained in \cite{Shomali2018,2Shomali2017}. The frequency interval from zero to $\omega_{max}$, driven for each dispersion curve, is divided to N$_{\rm int}$=1000. In this paper, the discrete frequency interval, $\Delta\omega_{i}$, is taken to be 5.917$\times$10$^{10}$, 4.94$\times$10$^{10}$, and 3.93$\times$10$^{10}$, respectively, for MoSi$_2$N$_4$, WSi$_2$N$_4$, and blue phosphorene. Also, the weighting factor W=$\frac{\textrm{N}_{\rm actual}}{\textrm{N}_{\rm prescribed}}$=1, is considered. Here, N$_{\rm prescribed}$ is the number of the stochastic samples which are actually initialized into the system. 
	\begin{table}[htbp]
		\caption{The coefficients of the fitted quadratic formula, $\omega$$_b$=c$_b$k$^{2}$+v$_b$k, for two studied low-dimensional materials.}
		\label{Tab1-tab1}
			\hspace{-0.8cm}
		\vspace{-0.5cm}
		\begin{center}
			\begin{small}
			
				\begin{tabular}{cccc}
					\hline
					2D Material    &  MoSi$_2$N$_4$    &
					WSi$_2$N$_4$  & BP    \\ \hline \hline
					c$_{LA}$(m$^2$/s)  & -4.252$\times$10$^{-7}$ & -3.392$\times$10$^{-7}$ & -5.9$\times$10$^{-7}$ \\ \hline
					c$_{TA}$(m$^2$/s) & -2.562$\times$10$^{-7}$ & -2.256$\times$10$^{-7}$ & -3.5$\times$10$^{-7}$ \\ \hline
					c$_{ZA}$(m$^2$/s) & -1.259$\times$10$^{-7}$ & -1.4$\times$10$^{-7}$  &  1.61$\times$10$^{-7}$\\ \hline
					v$_{LA}$(m/s) & 10031 & 8188 & 9668 \\ \hline
					v$_{TA}$(m/s) & 6982 & 5816  &  5897 \\ \hline
					v$_{ZA}$(m/s) & 5660 & 4886 &  896 \\ \hline
				\end{tabular}
			\end{small}
		\end{center}
	\end{table}

	During transport, phonons may experience collisions with other phonons, or they may scatter from the boundaries. In Monte Carlo simulation, the drift and scattering events are, subsequently, addressed individually \cite{Shomali2018,2Shomali2017}. Here, the boundaries are flat and impose the specular reflection. So, the phonon velocity along the boundary remains constant. Also, as the low-dimensional materials are assumed to be pure and perfect, the scatterings due to vacancies, dislocations, and impurities are neglected. In the temperature range of interest, 300 K-600 K, the phonon-phonon scattering is the most important mechanism for a suspended low-dimensional system. In the present study, the three-phonon Umklapp/normal scattering and the scattering from boundaries are the mechanisms that are considered. Although, the simplified description of the scattering dynamics via the relaxation-time approximation, cannot thoroughly describe the heat transport at low temperatures \cite{Cepellotti2017}, we use this approximation due to dealing with almost high temperature  \cite{Mazumder2001}. The Umklapp phonon-phonon scattering rate is calculated through $\tau^{-1}_{b,U}(\omega)=\frac{\hbar \gamma ^{2}_{b}}{\bar{M} \Theta_{b} v^{2}_{s,b}} \omega^{2} T e^{-\Theta_b/3T}$ \cite{THLiu2015}. Where, v$_{s,b}$, $\bar{M}$, and $\Theta$ are, respectively, the sound velocity for branch b, the average atomic mass, and the Debye temperature. The first part of the equation is the standard Umklapp interaction strength, and the exponential term presents the contribution from the redistribution by the N processes. The parameters $\Theta$, $\bar{M}$ and the Gr\"{u}neissen parameters, $\gamma$, used in this study, are displayed in the table \ref{Tab2-BP}.
	
	\begin{table}[htbp]
		\caption{The thermal characteristics applied for calculating relaxation time.  \newline}
		\label{Tab2-BP}
				\vspace{-0.5cm}

		\centering
		\begin{small}
			\hspace*{-0.4cm}
			\begin{tabular}{cccc}
				\hline
				2D Material &  MoSi$_2$N$_4$    &
				WSi$_2$N$_4$   & Blue phosphorene \\ \hline \hline
				$\Theta$  & 335.33 & 433.33  & 550 \\ \hline
				$\bar{M}$(e$^{-27}$ kg) & 29.74  & 42.29 & 51.4102 \\ \hline
				$\gamma$ &  0.41 & 0.4916 & 1.05 \\ \hline
			\end{tabular}
		\end{small}
	\end{table}
	
	As the higher order phonon scatterings get important at temperatures much larger than the Debye temperature and the operating temperature of most nano-electronic devices, they are not considered. Using the Monte Carlo technique, the probability of scattering a phonon with the rate of $\tau_b$ during the time t and t+$\Delta$t is obtained via, $\textrm{P}_{scat}=1-\exp(\frac{-\Delta t}{\tau})$. The scattering occurs if the probability of scattering, P$_{scat}$, is greater than a chosen random number R$_{scat}$. The frequency, branch, and direction of the phonons will be re-sampled from the cumulative density function \cite{Shomali2017}. We have chosen the time step, $\Delta$t, to be smaller than the minimum scattering rate of the phonons. The distribution function after scattering, F$_{scat}$ is modified by the probability of scattering, $\textrm{F}_{scat}(\tilde{T})=\frac{\sum_{j=1}^{i}\textrm{N}_j(\tilde{T}) \ \times \textrm{P}_{scat,j}}{\sum_{j=1}^{N_{b}}\textrm{N}_j(\tilde{T}) \ \times \textrm{P}_{scat,j}}$. This is done while the rate of the formation of phonons should be equal to its rate of destruction. More details can be found in \cite{Mittal2011,Shomali2017}.
	
	\section{Numerical method}
	\label{Numerical method}
	
	The frequency range for LA, TA, and ZA branches is divided into 1000 intervals. The scattering rates are obtained for the phonons with all possible 1000 frequencies in every branch. Accordingly, the phonon relaxation time is calculated by inversion of the scattering rate. Moreover, the velocities of all phonons with 1000 different frequencies are found for LA, TA, and ZA branches. By dividing the mesh size by the velocities, phonon traveling time is obtained. The minimum value of the calculated time is the considered time step. As a consequence, the time steps of 3.47$\times$10$^{-13}$, 2.07$\times$10$^{-13}$, and 3.46$\times$10$^{-13}$ are obtained, respectively, for  MoSi$_2$N$_4$, WSi$_2$N$_4$, and blue phosphorene. Mesh-independency test has also been carried out for every simulated channel. Throughout the entire time of the simulation, using a uniform mesh size of 200$\times$200 in XY-plane is found to be perfect for obtaining the mesh-independent plots. Also, the heat source zone is located in the center of the channel, in a cube of 30$\times$200$\times$0.53 nm$^{3}$. This term is simulated as a source that heats the desired zone by releasing the phonons into it. The methodology is completely described in \cite{Wong2011,Wong2014}.
	
	\section{Results and Discussions}
	\label{Results}
	
	In this section, the results for the temperature profile and the maximum temperature reached in the newly proposed FET with MA$_2$Z$_4$ channels are presented. The results are compared with the previously suggested FET with a thermally appropriate blue phosphorene channel \cite{Shomali2018}. The simulation situation is close to what exists in an actual FET. The simulated nano-devices are under the effect of the self-generated heat, caused mainly by the Joule-Heating process, during the first 200 ps of the computation time. Then, for the next 200 ps, the low-dimensional FETs are cooled. 
	
	\begin{table}[htbp]
		
		\vspace{-0.35cm}
		\begin{center}
			\begin{small}

				\begin{tabular}{|c|c|c|c|c|c|c|}
					\hline
					
					\multicolumn{4}{ |c| } {\cellcolor{pink!10}(a) MoSi$_2$N$_4$} \\
					\hline			t (ps) & LA (\%)&TA (\%)&ZA (\%) \\
					\hline 
					2 &  {\cellcolor{cyan!25}40} &  {\cellcolor{cyan!7}25}& {\cellcolor{cyan!15}35} \\
					\hline
					10 &  {\cellcolor{cyan!25}62} & {\cellcolor{cyan!7}14} & {\cellcolor{cyan!15}24} \\
					\hline
					50 &  {\cellcolor{cyan!25}66} &{\cellcolor{cyan!7}15} & {\cellcolor{cyan!15}19} \\
					\hline
					70 &  {\cellcolor{cyan!25}70} & {\cellcolor{cyan!7}11}& {\cellcolor{cyan!15}19} \\
					\hline
					100 &  {\cellcolor{cyan!25}72} & {\cellcolor{cyan!7}10} & {\cellcolor{cyan!15}18} \\
					\hline
					150 &  {\cellcolor{cyan!25}70} & {\cellcolor{cyan!7}13}& {\cellcolor{cyan!15}17} \\
					\hline
					200 &  {\cellcolor{cyan!25}70} & {\cellcolor{cyan!7}13}& {\cellcolor{cyan!15}17} \\
					\hline
					210 &  {\cellcolor{cyan!25}80} & {\cellcolor{cyan!7}8}& {\cellcolor{cyan!15}12} \\
					\hline
					240 &  {\cellcolor{cyan!25}71} & {\cellcolor{cyan!7}14}& {\cellcolor{cyan!15}15} \\
					\hline
					250 &  {\cellcolor{cyan!25}47} &{\cellcolor{cyan!15}28} & {\cellcolor{cyan!7}25} \\
					\hline
					300 &  {\cellcolor{cyan!25}44} & {\cellcolor{cyan!15}32}& {\cellcolor{cyan!7}24} \\
					\hline
					350 &  {\cellcolor{cyan!25}40} &{\cellcolor{cyan!15}34} & {\cellcolor{cyan!7}26} \\
					\hline
					400 &  {\cellcolor{cyan!25}42} & {\cellcolor{cyan!15}32}& {\cellcolor{cyan!7}26} \\
					\hline \hline
					
					\multicolumn{4}{ |c| } {\cellcolor{pink!10}(b) WSi$_2$N$_4$}\\
					\hline 
					
					t (ps) &  LA (\%)&TA (\%)&ZA (\%)\\
					\hline
					2 & {\cellcolor{RoyalBlue!7}30} & {\cellcolor{RoyalBlue!15}32}& {\cellcolor{RoyalBlue!25}38}  \\
					\hline
					10 & {\cellcolor{RoyalBlue!25}50} &  {\cellcolor{RoyalBlue!7}24} & {\cellcolor{RoyalBlue!15}26} \\
					\hline
					50 & {\cellcolor{RoyalBlue!25}52} & {\cellcolor{RoyalBlue!7}23}& {\cellcolor{RoyalBlue!15}25} \\
					\hline
					70 & {\cellcolor{RoyalBlue!25}51} &{\cellcolor{RoyalBlue!15}25} & {\cellcolor{RoyalBlue!7}24} \\
					\hline
					100 & {\cellcolor{RoyalBlue!25}53} & {\cellcolor{RoyalBlue!15}24}& {\cellcolor{RoyalBlue!7}23} \\
					\hline
					150 & {\cellcolor{RoyalBlue!25}54} &{\cellcolor{RoyalBlue!15}23} & {\cellcolor{RoyalBlue!15}23} \\
					\hline
					200 & {\cellcolor{RoyalBlue!25}54} & {\cellcolor{RoyalBlue!15}25}& {\cellcolor{RoyalBlue!7}21} \\
					\hline
					210 & {\cellcolor{RoyalBlue!25}56} & {\cellcolor{RoyalBlue!15}23}& {\cellcolor{RoyalBlue!7}21} \\
					\hline
					240 & {\cellcolor{RoyalBlue!15}44}  &{\cellcolor{RoyalBlue!25}33} & {\cellcolor{RoyalBlue!7}23} \\
					\hline
					250 & {\cellcolor{RoyalBlue!15}34}  &{\cellcolor{RoyalBlue!25}38} & {\cellcolor{RoyalBlue!7}26} \\
					\hline
					300 & {\cellcolor{RoyalBlue!15}32} & {\cellcolor{RoyalBlue!25}41}& {\cellcolor{RoyalBlue!7}27} \\
					\hline
					350 & {\cellcolor{RoyalBlue!7}25}  &{\cellcolor{RoyalBlue!25}45} & {\cellcolor{RoyalBlue!15}30} \\
					\hline
					400 & {\cellcolor{RoyalBlue!7}24} & {\cellcolor{RoyalBlue!25}46}& {\cellcolor{RoyalBlue!15}30} \\
					\hline
					\hline
					
					\multicolumn{4}{ |c| } {\cellcolor{pink!10} (c) Blue Phosphorene}\\
					\hline

					t (ps) & LA (\%)&TA (\%)&ZA (\%)\\
					\hline
					2 & {\cellcolor{blue!7}24} & {\cellcolor{blue!15}25} & {\cellcolor{blue!25}51}  \\
					\hline
					10 & {\cellcolor{blue!15}32} & {\cellcolor{blue!7}20} &  {\cellcolor{blue!25}48} \\
					\hline
					50 & {\cellcolor{blue!15}33} & {\cellcolor{blue!7}22} &  {\cellcolor{blue!25}45}  \\
					\hline
					70 & {\cellcolor{blue!15}35} & {\cellcolor{blue!7}20}&  {\cellcolor{blue!25}45} \\
					\hline
					100 & {\cellcolor{blue!7}28} &{\cellcolor{blue!15}30} &  {\cellcolor{blue!25}42} \\
					\hline
					150 & {\cellcolor{blue!15}34} & {\cellcolor{blue!7}18}&  {\cellcolor{blue!25}48} \\
					\hline
					200 & {\cellcolor{blue!15}35} & {\cellcolor{blue!7}17} &  {\cellcolor{blue!25}48} \\
					\hline
					210 & {\cellcolor{blue!7}32} & {\cellcolor{blue!15}34} &  {\cellcolor{blue!25}34} \\
					\hline
					240 & {\cellcolor{blue!7}30} & {\cellcolor{blue!15}34} &  {\cellcolor{blue!25}36} \\
					\hline
					250 & {\cellcolor{blue!7}23} & {\cellcolor{blue!15}38}&  {\cellcolor{blue!25}39} \\
					\hline
					300 & {\cellcolor{blue!7}21} &{\cellcolor{blue!15}24} &  {\cellcolor{blue!25}55} \\
					\hline
					350 & {\cellcolor{blue!7}14} &{\cellcolor{blue!15}24} &  {\cellcolor{blue!25}55} \\
					\hline
					400 & {\cellcolor{blue!7}10} &  {\cellcolor{blue!15}29}&   {\cellcolor{blue!25}61} \\
					\hline
					
				\end{tabular}
			\end{small}
		\end{center}
		
		\vspace{-0.5cm}
		\hspace{-1.7cm}	\caption{\label{branches}The percentage of the number of LA, TA, and ZA phonons which contribute to the heat transfer when the channel is made of (a) MoSi$_2$N$_4$, (b) WSi$_2$N$_4$, and (c) Blue phosphorene. The set of colors (light to dark) demonstrates the least to a maximum contribution of branches for each channel.}
	\end{table}
	
	\begin{table*}[htbp]
		\caption{The percentage of the frequency range for each phonon branch of (a) MoSi$_2$N$_4$, (b) WSi$_2$N$_4$, and (c) Blue phosphorene. Seriation of colors (light to dark) presents a low to high-value frequency range. All the contributions are given in percent, $\%$}
		\label{frequency}
		\begin{center}
			\begin{small}
				\vspace{-10pt}
				
				\begin{tabular}{|c|c|c|c|c|c|c|c|c|c|c|c|c|}
					
					\hline  
					\hhline{~~-----------}
					\hhline{~~-----------}	\rowfont{\huge}  \multicolumn{2}{ |c| } {\cellcolor{pink!10}(a) MoSi$_2$N$_4$}  & \multicolumn{11}{ !{\vrule width 0.5pt}c!{\vrule width 0.5pt}} {\cellcolor{lime!10} t (ps)}  \\  \hhline{--~~~~~~~~~~~} 
					{\cellcolor{lime!10}Mode} &\multicolumn{1}{c!{\vrule width 0.1pt} }{\cellcolor{lime!10}Frequency range (Hz)} &\multicolumn{1}{ !{\vrule width 0.5pt}c!} {2} &10 &50 &70 &100 &150 &200 &250 &300 &350 &400   \\  	 \hhline{--~~~~~~~~~~~}	 \hhline{-------------}  \hhline{-------------}
					\multirow{3}{*}{LA} &\ 0$\le\omega\le$4.66E13 ($\omega_{LA,TA,ZA}$) &{\cellcolor{cyan!7}2.1} & {\cellcolor{cyan!7}1}& {\cellcolor{cyan!7}0.7} &{\cellcolor{cyan!7}0.6}& {\cellcolor{cyan!7}0.2}&{\cellcolor{cyan!7}0.5}& {\cellcolor{cyan!7}0.4}&{\cellcolor{cyan!7}4.4}&{\cellcolor{cyan!7}1.3}&{\cellcolor{cyan!7}3}&{\cellcolor{cyan!7}3.2}\\
					\cline{2-13} &4.66E13$\le\omega\le$4.80E13 ($\omega_{LA,ZA}$)&{\cellcolor{cyan!15}0.2} &{\cellcolor{cyan!15}0.2} &{\cellcolor{cyan!15}0.1} &{\cellcolor{cyan!15}0.2}& {\cellcolor{cyan!15}0.1}&{\cellcolor{cyan!15}0.1}&{\cellcolor{cyan!15}0.1}&{\cellcolor{cyan!15}0.4}&{\cellcolor{cyan!15}0.4}&{\cellcolor{cyan!15}0.5}&{\cellcolor{cyan!15}0.6}\\
					\cline{2-13}& 4.80E13$\le\omega\le$5.917E13 ($\omega_{LA}$) &{\cellcolor{cyan!25}97.7} &{\cellcolor{cyan!25}98.8} & {\cellcolor{cyan!25}99.2} &{\cellcolor{cyan!25}99.2}& {\cellcolor{cyan!25}99.7}&{\cellcolor{cyan!25}99.4}& {\cellcolor{cyan!25}99.5}&{\cellcolor{cyan!25}95.2}&{\cellcolor{cyan!25}98.3}&{\cellcolor{cyan!25}96.5}&{\cellcolor{cyan!25}96.2}\\
					\hline
					\hhline{-------------}	\hhline{-------------}
					\multirow{2}{*}{ZA} &\ 0$\le\omega\le$4.66E13 ($\omega_{LA,TA,ZA}$) & {\cellcolor{cyan!7}99.8} & {\cellcolor{cyan!7}99.7}& {\cellcolor{cyan!7}99.7} & {\cellcolor{cyan!7}99.7}& {\cellcolor{cyan!7}99.6}& {\cellcolor{cyan!7}99.9}& {\cellcolor{cyan!7}99.9}&{\cellcolor{cyan!7}99.7}&{\cellcolor{cyan!7}99.6}&{\cellcolor{cyan!7}99.7}&{\cellcolor{cyan!7}99.7}\\
					\cline{2-13} &4.66E13$\le\omega\le$4.80E13 ($\omega_{LA,ZA}$)& {\cellcolor{cyan!15}0.2} & {\cellcolor{cyan!15}0.3} & {\cellcolor{cyan!15}0.3} & {\cellcolor{cyan!15}0.3}& {\cellcolor{cyan!15}0.4}&{\cellcolor{cyan!15}0.1}& {\cellcolor{cyan!15}0.1}&{\cellcolor{cyan!15}0.3}&{\cellcolor{cyan!15}0.4}&{\cellcolor{cyan!15}0.3}&{\cellcolor{cyan!15}0.3}\\
					\hline
					\hhline{-------------}	\hhline{-------------}
					\multirow{1}{*}{TA} &\ 0$\le\omega\le$4.66E13 ($\omega_{LA,TA,ZA}$) & {\cellcolor{cyan!7}100} & {\cellcolor{cyan!7}100}& {\cellcolor{cyan!7}100} &{\cellcolor{cyan!7}100}&{\cellcolor{cyan!7}100}&{\cellcolor{cyan!7}100}&{\cellcolor{cyan!7}100}&{\cellcolor{cyan!7}100}&{\cellcolor{cyan!7}100}&{\cellcolor{cyan!7}100}&{\cellcolor{cyan!7}100}\\
					\cline{1-13} 	\hline 	 \hline
					
					\hline  
					\hhline{~~-----------}
					\hhline{~~-----------}
					\rowfont{\huge} \multicolumn{2}{|c!{\vrule width 0.1pt} } {\cellcolor{pink!10}\ \ \ (b) WSi$_2$N$_4$}  & \multicolumn{11}{ !{\vrule width 0.5pt}c!{\vrule width 0.5pt}} {\cellcolor{lime!10} t (ps)} \\  
					
					\hhline{--~~~~~~~~~~~} 	{\cellcolor{lime!10}Mode}  &\multicolumn{1}{c!{\vrule width 0.1pt} }{\cellcolor{lime!10}Frequency range (Hz)} & \multicolumn{1}{ !{\vrule width 0.5pt}c!} {2} &\multicolumn{1}{ |c| } {10} &\multicolumn{1}{ c| } {50} &\multicolumn{1}{ c| } {70} &\multicolumn{1}{ c| } {100} &\multicolumn{1}{ c| } {150} &\multicolumn{1}{ c| } {200} &\multicolumn{1}{ c| } {250} &\multicolumn{1}{ c| } {300} &\multicolumn{1}{ c| } {350} & \multicolumn{1}{c| } {400}   \\  	 \hhline{--~~~~~~~~~~~}	 \hhline{-------------}  \hhline{-------------}

					\multirow{3}{*}{LA} &\multicolumn{1}{ c| } { 0$\le\omega\le$3.74E13 ($\omega_{LA,TA,ZA}$)} & \multicolumn{1}{c|} {\cellcolor{RoyalBlue!7}0.62} & \multicolumn{1}{c|} {\cellcolor{RoyalBlue!7}1.7} & \multicolumn{1}{c|} {\cellcolor{RoyalBlue!7}1}&\multicolumn{1}{c|} {\cellcolor{RoyalBlue!7}0.9}&\multicolumn{1}{c|} {\cellcolor{RoyalBlue!7}1.2}&\multicolumn{1}{c|} {\cellcolor{RoyalBlue!7}1.1}&\multicolumn{1}{c|} {\cellcolor{RoyalBlue!7}1.15}&\multicolumn{1}{c|} {\cellcolor{RoyalBlue!7}7}&\multicolumn{1}{c|} {\cellcolor{RoyalBlue!7}4}&\multicolumn{1}{c|} {\cellcolor{RoyalBlue!7}6}&\multicolumn{1}{c!{\vrule width 0.1pt} } {\cellcolor{RoyalBlue!7}1.6}\\ 
					&\multicolumn{1}{ c| } {3.74E13$\le\omega\le$3.91E13 ($\omega_{LA,ZA}$)}&  {\cellcolor{RoyalBlue!15}0.18} &  {\cellcolor{RoyalBlue!15}0.4} &  {\cellcolor{RoyalBlue!15}0.3}&  {\cellcolor{RoyalBlue!15}0.3}&  {\cellcolor{RoyalBlue!15}0.4}&  {\cellcolor{RoyalBlue!15}0.1}&  {\cellcolor{RoyalBlue!15}0.25}&  {\cellcolor{RoyalBlue!15}1}&  {\cellcolor{RoyalBlue!15}1}&  {\cellcolor{RoyalBlue!15}1}&  {\cellcolor{RoyalBlue!15}0.6}\\
					\hhline{~------------}  & \multicolumn{1}{ c| } {3.74E13$\le\omega\le$4.94E13 ($\omega_{LA}$)} &  {\cellcolor{RoyalBlue!25}98.2} &  {\cellcolor{RoyalBlue!25}97.9} &  {\cellcolor{RoyalBlue!25}98.7}&  {\cellcolor{RoyalBlue!25}98.8}&  {\cellcolor{RoyalBlue!25}98.6}&  {\cellcolor{RoyalBlue!25}98.8}&  {\cellcolor{RoyalBlue!25}98.6}&  {\cellcolor{RoyalBlue!25}92}&  {\cellcolor{RoyalBlue!25}95}&  {\cellcolor{RoyalBlue!25}93}&  {\cellcolor{RoyalBlue!25}97.8}\\
					\hline
					\hhline{-------------}	\hhline{-------------}
					&\multicolumn{1}{ c| } {\ 0$\le\omega\le$3.74E13 ($\omega_{LA,TA,ZA}$)} &  {\cellcolor{RoyalBlue!7}84.4} & {\cellcolor{RoyalBlue!7}85.8} & {\cellcolor{RoyalBlue!7}86.1}  &{\cellcolor{RoyalBlue!7}86.2} & {\cellcolor{RoyalBlue!7}85.3}& {\cellcolor{RoyalBlue!7}85.8}& {\cellcolor{RoyalBlue!7}85.1}&{\cellcolor{RoyalBlue!7}78.6}&{\cellcolor{RoyalBlue!7}83.2}&{\cellcolor{RoyalBlue!7}77.85}& {\cellcolor{RoyalBlue!7}76.5}\\
					\hhline{~------------} 		\multirow{-2}{*}{ZA} &\multicolumn{1}{ c| } {3.74E13$\le\omega\le$3.91E13 ($\omega_{LA,ZA}$)}& {\cellcolor{RoyalBlue!15}15.6} & {\cellcolor{RoyalBlue!15}14.2} &  {\cellcolor{RoyalBlue!15}13.9} & {\cellcolor{RoyalBlue!15}13.8} & {\cellcolor{RoyalBlue!15}14.7}& {\cellcolor{RoyalBlue!15}14.2}& {\cellcolor{RoyalBlue!15}14.9}&{\cellcolor{RoyalBlue!15}21.4}&{\cellcolor{RoyalBlue!15}16.8}&{\cellcolor{RoyalBlue!15}22.15}& {\cellcolor{RoyalBlue!15}23.5}\\
					\hline
					\hhline{-------------}	\hhline{-------------}
					
					\multirow{1}{*}{TA} &\multicolumn{1}{ c| } {0$\le\omega\le$3.74E13 ($\omega_{TA}$)} & {\cellcolor{RoyalBlue!7}100} & {\cellcolor{RoyalBlue!7}100}& {\cellcolor{RoyalBlue!7}100} &{\cellcolor{RoyalBlue!7}100}&{\cellcolor{RoyalBlue!7}100}&{\cellcolor{RoyalBlue!7}100}&{\cellcolor{RoyalBlue!7}100}&{\cellcolor{RoyalBlue!7}100}&{\cellcolor{RoyalBlue!7}100}&{\cellcolor{RoyalBlue!7}100}&{\cellcolor{RoyalBlue!7}100}\\
					\hline \hline
					
					\rowfont{\huge}		\hhline{~~-----------}
					\hhline{~~-----------}	 \multicolumn{2}{|c|} {\cellcolor{pink!10}(c) Blue phosphorene}  &\multicolumn{11}{ !{\vrule width 0.5pt}c!{\vrule width 0.5pt}} {\cellcolor{lime!10} t (ps)}   \\ 		\hhline{--~~~~~~~~~~~} 
					{\cellcolor{lime!10}Mode} &\multicolumn{1}{c!{\vrule width 0.1pt} }{\cellcolor{lime!10}Frequency range (Hz)} &\multicolumn{1}{ !{\vrule width 0.5pt}c!} {2} &10 &50 &70 &100 &150 &200 &250 &300 &350 &400  \\  	 \hhline{--~~~~~~~~~~~}	 \hhline{-------------}  \hhline{-------------}
					\multirow{3}{*}{LA} &\ 0$\le\omega\le$2.48E13($\omega_{LA,TA,ZA}$) & {\cellcolor{blue!7}1.3} &  {\cellcolor{blue!7}0.6} &  {\cellcolor{blue!7}0.4} &  {\cellcolor{blue!7}0.51}&  {\cellcolor{blue!7}0.5} & {\cellcolor{blue!7}0.4}& {\cellcolor{blue!7}0.5}& {\cellcolor{blue!7}0.3}& {\cellcolor{blue!7}1.3}& {\cellcolor{blue!7}1.3}& {\cellcolor{blue!7}1.1}\\
					\cline{2-13} &2.48E13$\le\omega\le$2.94E13($\omega_{LA,ZA}$)&  {\cellcolor{blue!15}1}& {\cellcolor{blue!15}0.7} & {\cellcolor{blue!15}1.0} & {\cellcolor{blue!15}0.4} &{\cellcolor{blue!15}0.49} & {\cellcolor{blue!15}0.48}&{\cellcolor{blue!15}0.3}&{\cellcolor{blue!15}0.5}&{\cellcolor{blue!15}0.3}&{\cellcolor{blue!15}1.2}&{\cellcolor{blue!15}1.5}\\
					\cline{2-13}& 2.94E13$\le\omega\le$3.93E13($\omega_{LA}$) & {\cellcolor{blue!25}97.7} & {\cellcolor{blue!25}98.7} &  {\cellcolor{blue!25}99.1} & {\cellcolor{blue!25}99.2}&  {\cellcolor{blue!25}99.3}& {\cellcolor{blue!25}99}& {\cellcolor{blue!25}99.4}& {\cellcolor{blue!25}97.5}& {\cellcolor{blue!25}97.2}& {\cellcolor{blue!25}97.8}& {\cellcolor{blue!25}97}\\
					\hline
					\hhline{-------------}	\hhline{-------------}
					\multirow{2}{*}{ZA} &\ 0$\le\omega\le$2.48E13($\omega_{LA,TA,ZA}$) & {\cellcolor{blue!7}40} & {\cellcolor{blue!7}92.5}& {\cellcolor{blue!7}93.8} &{\cellcolor{blue!7}94.6}&{\cellcolor{blue!7}95.3}&{\cellcolor{blue!7}95.1}&{\cellcolor{blue!7}95.4}&{\cellcolor{blue!7}91.7}&{\cellcolor{blue!7}91.4}&{\cellcolor{blue!7}89.6}&{\cellcolor{blue!7}91.1}\\
					\cline{2-13} &2.48E13$\le\omega\le$2.94E13($\omega_{LA,ZA}$)& {\cellcolor{blue!15}62} & {\cellcolor{blue!15}7.5} & {\cellcolor{blue!15}6.2} &{\cellcolor{blue!15}5.4}&{\cellcolor{blue!15}4.7}&{\cellcolor{blue!15}4.9}&{\cellcolor{blue!15}4.6}&{\cellcolor{blue!15}8.3}&{\cellcolor{blue!15}8.6}&{\cellcolor{blue!15}10.4}&{\cellcolor{blue!15}8.9}\\
					
					\hline
					\hhline{-------------}	\hhline{-------------}
					\multirow{1}{*}{TA} &\ 0$\le\omega\le$2.48E13($\omega_{LA,TA,ZA}$) & {\cellcolor{blue!7}100} & {\cellcolor{blue!7}100}& {\cellcolor{blue!7}100} &{\cellcolor{blue!7}100}&{\cellcolor{blue!7}100}&{\cellcolor{blue!7}100}&{\cellcolor{blue!7}100}&{\cellcolor{blue!7}100}&{\cellcolor{blue!7}100}&{\cellcolor{blue!7}100}&{\cellcolor{blue!7}100}\\
					\cline{1-13} 	\hline 	 
					
				\end{tabular}
			\end{small}
		\end{center}
	\end{table*}
	Implementation of Monte Carlo simulation of phonon Boltzmann equation in low dimensional material, the types of the phonons which contribute to the heat transport are calculated and shown in the table. \ref{branches}. As it is presented in the table. \ref{branches}(a), most of the phonons involved in heat transport during the 200 ps of heating for the material MoSi$_2$N$_4$ are LA phonons. More particularly, in the early seconds, the contributions of all three phonon branches are nearly the same. Shortly after, the percentage of LA phonons increases abruptly such that it reaches 70$\%$ at 200 ps. At the same time, the number of ZA phonons decreases from 35$\%$ of the heat carriers at two ps to 17$\%$ at 200 ps. When the heat generation zone is turned off, and the cooling starts, the procedure becomes different, and the arrangement of the phonons changes. The phonons in the LA branch are lessened to 40$\%$, but instead, the number of phonons in TA mode grows up to three times the one at heating time. As a result, the total number of fast carriers remains around 70-80$\%$. Notwithstanding, slow flexural ZA phonons' participation is much lower and stays around 20-30 percent within the cooling and heating period. 
	
	Accordingly, an almost similar situation comes about for the WSi$_2$N$_4$ channel. While the participation of ZA phonons is notable for the few first picoseconds, it starts decreasing as time passes. For example, at t=200 ps, when the heating is turned off, the number of ZA phonons taking part in heat transfer is halved. This is true while the contribution of LA phonons during the heating is doubled. On the other hand, when the heat generation is turned off, and throughout the 200 ps of cooling, the leading heat carriers transform to TA phonons. In summary, within the whole time of simulation, nearly 70-80$\%$ of the heat transport is carried out via the fast high- and low-frequency LA and TA phonons. When the transistor is cooled between 200-400 ps, the number of LA phonons is nearly halved, while the TA phonons are duplicated. However, in similarity to the transistor with MoSi$_2$N$_4$ channel, the total number of fast phonons is still much more significant than the slow ZA phonons. While regarding to MoSi$_2$N$_4$, much more contribution belongs to the lower frequency TA mode. On the whole, the dominant heat carriers in MA$_2$Z$_4$ type materials are LA and TA phonons, while during the heating, the portion of LA mode for MoSi$_2$N$_4$ is up to 20$\%$ larger than that of WSi$_2$N$_4$ channel. In more detail, during the simulation, the absolute value of the difference between the number of LA and TA phonons, are between 2-30$\%$ and 6-62$\%$, respectively, for WSi$_2$N$_4$ and MoSi$_2$N$_4$. Consequently, high-frequency phonons are significantly more in MoSi$_2$N$_4$ material. 

	In the next step, the phonon analysis for the blue phosphorene channel is performed. As is seen in the table. \ref{branches}(c), BP exhibits the reverse trend. In more detail, almost during the entire simulation, half of the heat carriers are the slow lower frequency phonons, in flexural direction, ZA mode. Meanwhile, the contribution of LA and TA phonons are approximately the same and much less than the flexural ones in such a way that the contribution of both is about the single contribution of the ZA phonons. This behavior is held during the heating and cooling processes.
	
    Furthermore, in this study, the frequency of the heat carriers from various branches contributing to the heat transfer are analyzed. The table \label{frequncy} demonstrates the frequency range for phonons in the LA, TA, and ZA branches of three materials of MoSi$_2$N$_4$, WSi$_2$N$_4$, and Blue phosphorene. The situation for MoSi$_2$N$_4$ is described in detail in the table. \ref{frequency}(a). It was previously shown that most of the heat carriers in MoSi$_2$N$_4$ were longitudinal acoustic phonons. As it is obvious, within the heating and cooling processes, more than 90$\%$ of the LA phonons, which are participated in heat transfer, always have a frequency between 3.74-4.94$\times$10$^{13}$ Hz, which is always higher than the frequency of any phonons in TA or ZA modes. So, the phonons taking part in heat transport are fast with high frequency, which compete with each other to specify the value of the maximum temperature. The table. \ref{frequency}(b) confirms the result for the material WSi$_2$N$_4$. The frequency of the LA phonons carrying heat is like what occurs in MoSi$_2$N$_4$, where between 93-98$\%$ of the LA phonons have frequencies larger than that of TA and ZA phonons. Differently in comparison to MoSi$_2$N$_4$, the number of TA phonons involved in the thermal transfer is up to two times larger. The TA phonons with the lowest frequency range in the phonon dispersion curve, are also fast and can leave the hotspot region promptly. In both MA$_2$Z$_4$ materials, while the contribution of slow ZA phonons in heat transport is low, 80-85$\%$ of them have a frequency less than the maximum frequency of TA modes. Here is where the velocity of ZA phonons and the lower frequency contend reversely for maximum temperature, as compared to fast high-frequency LA phonons of the material MA$_2$Z$_4$. 
	
	Although the combination of phonons and the type of the dominant phonons which participate in heat transport for Blue phosphorene is utterly different from that of MoSi$_2$N$_4$ and WSi$_2$N$_4$, the frequency range behavior for different types of the phonons is the same. As the table. \ref{frequency}(c) shows, when the contribution of the LA phonons is between 10 to 35 percent, these almost low-in-number phonons have a frequency larger than the maximum frequency of TA and ZA modes. Also most of the phonons in TA mode, have a frequency lower than ZA maximum frequency. Finally, the ZA phonons, which are the main contributions to the heat transfer in BP, are in the lowest frequency range of 0-2.48$\times$10$^{13}$ Hz. Here, likewise, the behavior of the ZA phonons in MoSi$_2$N$_4$ and WSi$_2$N$_4$ materials, the low velocity of ZA phonons which causes them to trap in hotspot, compete with their low-frequency to determine the maximum temperature. 
	
	\begin{figure}
	\centering
					\vspace{-3.12cm}
		\includegraphics[width=\columnwidth]{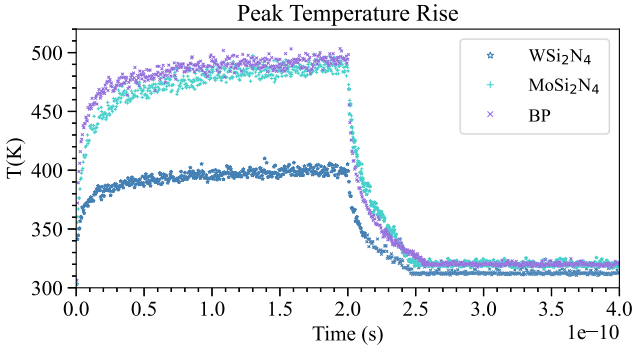}
				\vspace{-3.12cm}

		\caption{\label{Fig2}The maximum temperature versus time at the XY plane. $+$, star, and $\times$ markers, respectively, demonstrate the result for MoSi$_2$N$_4$, WSi$_2$N$_4$, and Blue phosphorene. The WSi$_2$N$_4$ channel has reached a lower peak temperature and can be proposed as the very thermally efficient candidate for silicon replacement.}
		\vspace{-13pt}
	\end{figure}
	As formerly mentioned, the reliability of transistors depends on the maximum temperature they reach \cite{Moore2014}. Hence, the peak temperature rise during the heating and cooling is analyzed in detail. In Fig. \ref{Fig2}, the behavior of maximum temperature for MoSi$_2$N$_4$, WSi$_2$N$_4$, and the test blue phosphorene channels versus the time is demonstrated. The figure presents the average result of ten complete runs of the simulation. As is seen, ignoring the MoSi$_2$N$_4$ material behavior, which reaches the maximum temperature and the steady state peak temperature in a longer time, the three materials nearly follow the same trend. For the first 200 ps, while heating is occurring, the temperature increases, and then for the second cooling 200 ps, the temperature decreases to the steady value. The WSi$_2$N$_4$ material presents the highest temperature, much less than that of MoSi$_2$N$_4$ and Blue phosphorene. In accordance with the previous heat carrier analysis demonstrated in the table. \ref{branches}, the dominant co-existence of the much faster LA and TA phonons is one of the reasons for the lower maximum temperature. The TA branch phonons, on one side, are fast to free themselves out of the hotspot, and on the other side, as the table. \ref{frequency} confirms, have frequencies that are comparable to the ZA phonons and not as large as that of the LA mode. As earlier stated, for the MoSi$_2$N$_4$ channel, most of the heat is transferred through LA phonons. The high velocity of these carriers lets them escape the hotspot, but their much higher frequency makes the temperature rise more. Thereby, the competition winner between the heat carrier velocity and the related frequency is responsible for the maximum reached temperature. As Fig. \ref{Fig2} establishes, for the WSi$_2$N$_4$ material, the velocity is almost high, and simultaneously, the frequency value is relatively low. Hence, both parameters lead to the very appropriate thermal condition and make the maximum temperature of the WSi$_2$N$_4$ very low in comparison to the other studied substances. This contrasts to the condition that the LA phonons create in MoSi$_2$N$_4$ channel with the desirable thermal and electrical properties. For the MoSi$_2$N$_4$, the effect of frequency overcomes the velocity of high-frequency LA phonons, which are primarily responsible for heat transfer phenomena. So, this makes MoSi$_2$N$_4$ substance by a large margin as the second stage with less peak temperature achieved. The obtained maximum temperature is, respectively, signifiently less and slightly higher than that of the WSi$_2$N$_4$ and blue phosphorene. For BP, the important TA phonons are fewer, and at the same time, the number of the low-velocity ZA phonons with a frequency comparable to TA ones is high and prevailing. So, both conditions are thermally worse than what exists in MoSi$_2$N$_4$ and WSi$_2$N$_4$ materials, and so this leads up to much higher temperature. Accordingly, the blue phosphorene, which was formerly nominated as the best choice through the candidate 2-D channels of Germanene, Silicene, MoS$_2$ and, Graphene \cite{Shomali2018}, shows the worst condition in comparison to MA$_2$Z$_4$ materials and specifically WSi$_2$N$_4$. Furthermore, the Fig. \ref{Fig2} presents that the material WSi$_2$N$_4$ attains a steady state at a temperature which is 10 K less than the steady state maximum temperature of the MoSi$_2$N$_4$ and Blue phosphorene. 
	
	\begin{figure}
		\centering
		\includegraphics[width=1.05\columnwidth]{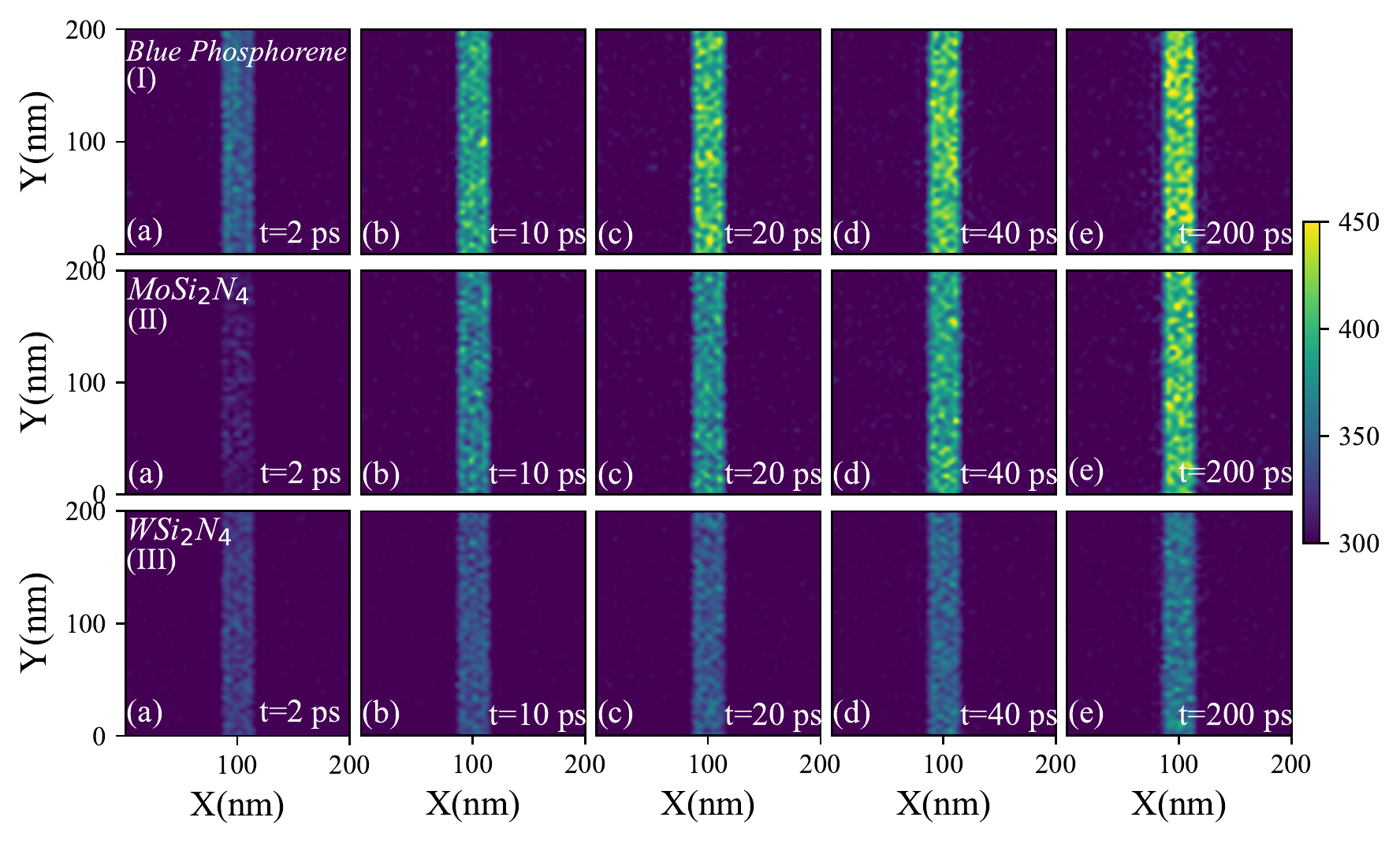}
		\caption{\label{Fig3} The behavior of the hotspot at XY plane when (a) t=2 ps, (b) t=10 ps, (c) t=20 ps, (d) t=40 ps, (e) t=200 ps. The upper (I), middle (II), and lower (III) plots, subsequently, present the situation for the materials blue phosphorene, MoSi$_2$N$_4$, and WSi$_2$N$_4$. The formed hotspot for WSi$_2$N$_4$ has a lower temperature during the heating and cooling.}
	\end{figure}
	The formation of hotspots at proposed transistors is shown in Fig. \ref{Fig3}-I(a-e). As is seen, immediately after turning on the transistor, the hot spots are formed. The FETs including, the stable thermally efficient monolayer 2D material of type MA$_2$Z$_4$, named WSi$_2$N$_4$, present hotspots with signifiently lower maximum temperature. Specifically, as is seen in Fig. \ref{Fig3} (e), the hotspot for the thermally isotropic material BP with thermal conductivity of 75 Wm$^{−1}$ K$^{−1}$ reaches the highest temperature at t=200 ps. At first 200 ps, the hot region contentiously produces phonons with high frequency, but due to the predominant ZA phonons, the heat dissipation is very low. So the phonons trap in the hotspot and cannot leave it. Accordingly, the temperature at hotspots increases much. For MoSi$_2$N$_4$ at t=2ps, when the heating has just turned on, the hotspot is cooler. The situation is detectable in Fig. \ref{Fig3}-II(a). This behavior can be attributed to the notable contribution of both TA and LA mode phonons relative to what exists in larger times. In other words, both LA phonons, as the dominant ones with 40$\%$ participation, and the TA phonons with almost 25$\%$ contribution, cause the lower temperature. As time passes, the contribution of LA phonons with high frequency abruptly increases while the contribution of TA phonons decreases to 10$\%$. In consequence, the high-frequency LA phonons are prevailing, and the hotspot gets much higher. Since the circumstance continues this way, the dominance of TA over LA, and also exclusively the high frequency of LA over its high velocity, persists. The situation and the phonon analysis during the whole time of heating in WSi$_2$N$_4$ are the same as what occurs in the first picoseconds of the MoSi$_2$N$_4$ heating. As a result, the cooler hotspots are formed during the entire simulation of heating in a WSi$_2$N$_4$ channel which are presented in Fig. \ref{Fig3}-III(a-e). To be more precise, the hotspot for the material WSi$_2$N$_4$ with a notable thermal conductivity of 224 Wm$^{−1}$ K$^{−1}$ reaches the lowest temperature, see Fig. \ref{Fig3}-III(e). 
	
	\begin{figure}
		\hspace{0.5cm}
		\centering
		\includegraphics[width=1.05\columnwidth]{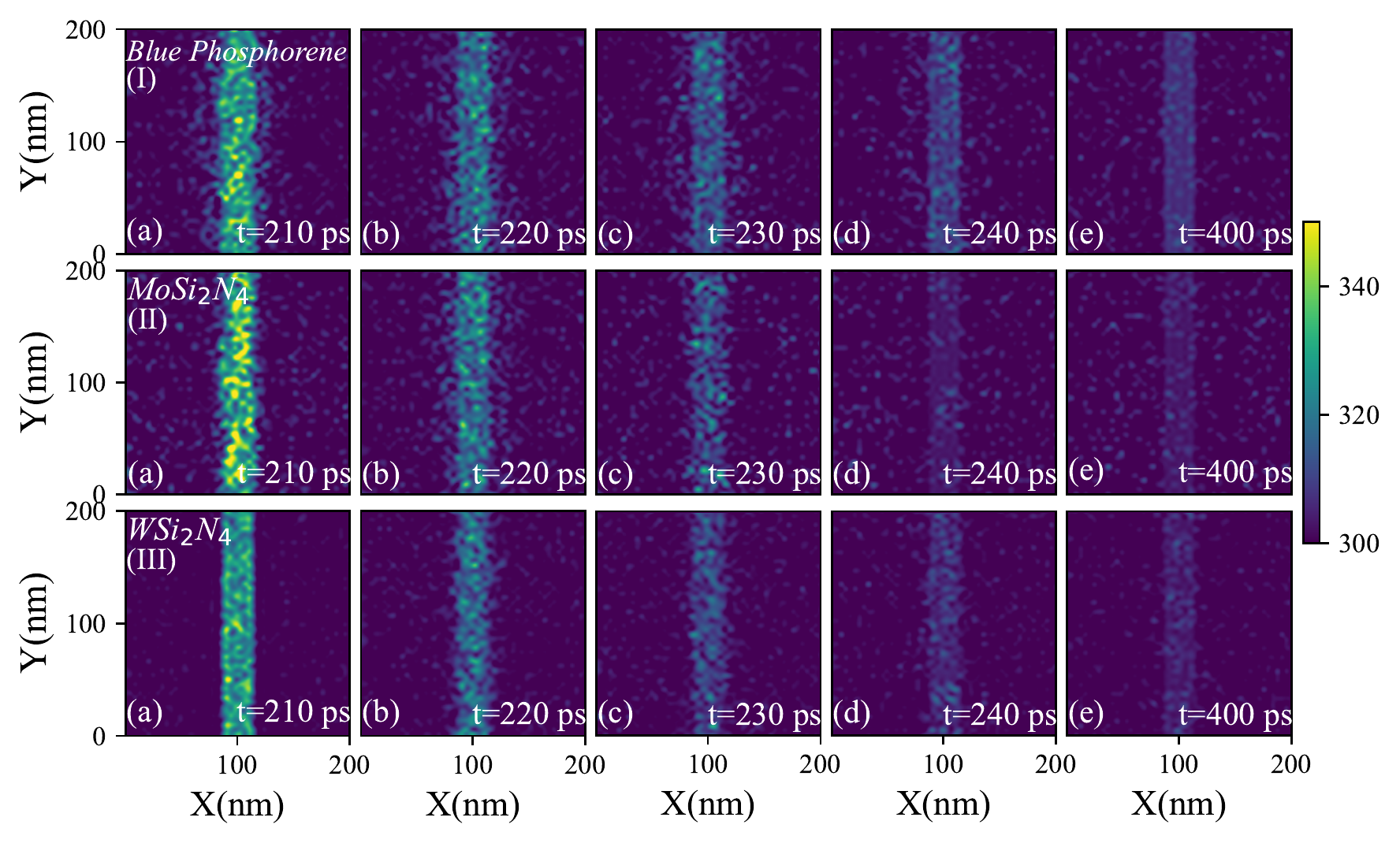}
		\caption{\label{Fig4} Hotspots at XY plane during the cooling when (a) t=210 ps, (b) t=220 ps, (c) t=230 ps, (d) t=240 ps, (e) t=400 ps for blue phosphorene (I), MoSi$_2$N$_4$ (II), and WSi$_2$N$_4$ (III). Note that the colorbars are different in Figs. \ref{Fig3} and \ref{Fig4}.}
	\end{figure}
	
	Moreover, as Fig. \ref{Fig4} suggests, the behavior of the hotspot during the cooling in all three studied channels, is conspicuous as well. When the heating is turned off, the hotpot starts to dissolve. It is established in Figs. \ref{Fig4}(I-III)(a-b) and \ref{Fig2}, that the condition for almost the first 30 ps of cooling is different. The circumstances can be justified while paying attention to the types of phonons which take part in heat transport. To make the situation clearer, the phonon analysis for t=210 ps will be given. As Figs. \ref{Fig4} (I-II)(a) confirm, at t=210 ps, the hotspot in MoSi$_2$N$_4$ is hotter than the one in BP. This is true while in t=200 ps, when the heating is switched off, BP shows the formed hotspot with a higher temperature. In consequence, although at t=200 ps, MoSi$_2$N$_4$ is with a cooler hotspot, the situation reverses at t=210 when the BP demonstrates a hotspot with a lower temperature. This finding confirms that the heat spreading process during the cooling for the BP channel is much more beneficial and efficient compared to heat propagation in MoSi$_2$N$_4$. Considering what is presented in the table. \ref{branches}, the status is vindicated. For MoSi$_2$N$_4$, when time passes t=200 ps, increasing the number of LA phonons and decreasing the TA phonons continues. In particular, at t=210 ps, the percentage of phonons participation is 80$\%$ and 8$\%$ for LAs and TAs. On the contrary, the presence of TA phonons in BP augments as the cooling is started. Explicitly, at t=210 ps, the presence of TA phonons in the hotspot is doubled relative to the time t=200 ps. Meanwhile, the LA and ZA phonons are reduced. As already indicated, the presence of TA phonons which are almost low energy particles, makes the energy lessens, and the heat spreading easier and consequently lowers the peak temperature rise. Also, despite that the contribution of slow and almost low energy ZA phonons for MoSi$_2$N$_4$ and BP are, respectively, 22$\%$ and 39$\%$. Nevertheless, it is evident that the presence of more TA phonons overcomes major ZA participation. This can also be assigned to less energy of TA phonons. Proportionately, as one expects, the heat propagation in blue phosphorene is strengthened during this period, and the hotspot cools down more. This trend remains within almost the first 30 ps of cooling time. At larger times, the contribution of high-energy LA phonons is reduced, while for MoSi$_2$N$_4$, the contribution of slow ZA phonons is also meager. On the other hand, for the channel BP, the number of LA phonons is also decreased, but simultaneously, the participation of slow ZA phonons has increased up to $60\%$. As ZA phonons are too slow relative to ZA and LA, the phonons leave the hotspot more leisurely. After t=230 ps, confirmed by Figs. \ref{Fig4} (I-II)(c-e), the effect of the number of ZA phonons on hotspot temperature conquers other items, and hence, in comparison, it is anticipated that after t=230 ps, the MoSi$_2$N$_4$ has cooler hotspot to the lack of ZA phonons. In other words, at larger times, the heat spread in two MA$_2$Z$_4$ materials is faster than the blue phosphorene, which results in a lower temperature steady state. Further, as Figs. \ref{Fig4} (a-e) manifest, the WSi$_2$N$_4$ channel leads the deportment similar to that of the MoSi$_2$N$_4$. Moreover, the time needed for the transistor to reach the steady state grows, respectively, for channels made of WSi$_2$N$_4$, MoSi$_2$N$_4$, and BP. In brief, Figs. \ref{Fig3} and \ref{Fig4} confirm that the MA$_2$Z$_4$  materials with shallow maximum temperatures are the most reliable materials for silicon replacement, such that they are even more preferable than the suggested Blue phosphorene with favorable thermal and electrical characteristics. Figs. \ref{Fig3} (a-c) affirm that the WSi$_2$N$_4$ from MA$_2$Z$_4$ family with very low maximum temperature is the most reliable material for silicon replacement, such that it is even more preferable than the suggested Blue phosphorene with favorable thermal and electrical characteristics.
	
	\begin{figure*}
		\vspace{-1.12cm}
		\centering
		\includegraphics[width=1.3\columnwidth]{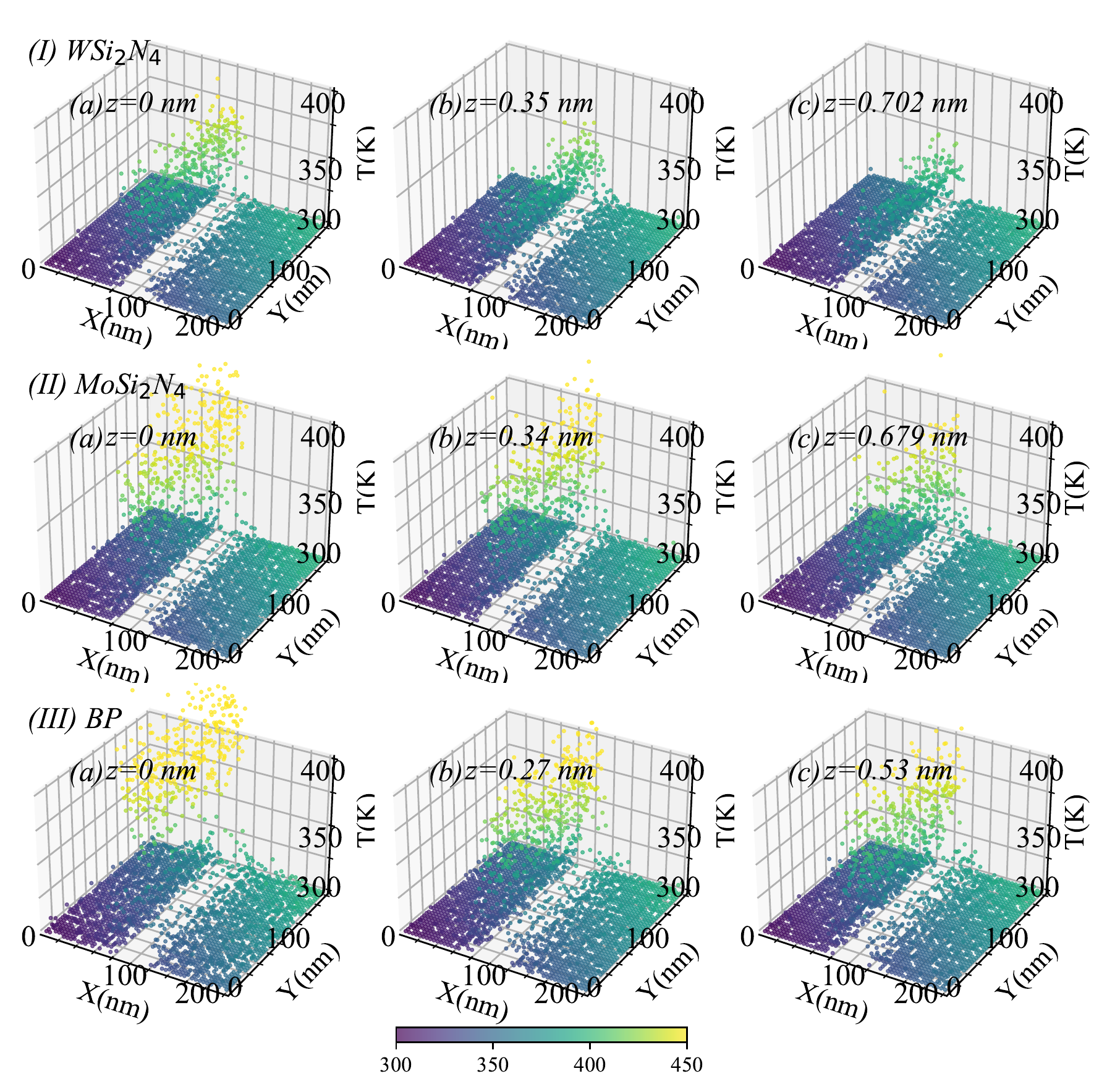}
		\caption{\label{Fig5}The temperature profile at XY plane for (I) WSi$_2$N$_4$, (II) MoSi$_2$N$_4$ and (III) blue phosphorene at (a) bottom boundary, z=0, (b) z equal to half the thickness, and (c) top boundary, z=thickness at t=200 ps when the heating is switched off.}
	\end{figure*}
	
	At last, the temperature profile at different positions of channels is discussed. The three materials of WSi$_2$N$_4$, WSi$_2$N$_4$, and BP have thicknesses of 0.679, 0.702, and 0.53. As already pointed, all boundaries have been considered adiabatic except the bottom boundary, which can exchange energy with the environment. The thicknesses of the materials are very low, so, some of the phonons reach the bottom boundary in almost the time that are created in the hotspot. However, until time t=200 ps, the heat generation zone creates the phonons consecutively and correspondingly keeps the temperature of the top boundary high (see Figs. \ref{Fig5} (I-III)(a)). At the same time, hot phonons arriving at the bottom boundary can leave the channel. As a result, like what is shown in Figs. \ref{Fig5} (I-III)(c), the hotspot at the bottom boundary is much cooler. Also, this is the case when one deals with the formed hotspot at z=d/2, with d being the channel thickness. In comparison to the top/bottom boundary, less/more phonons leave this region. Accordingly, as is seen in Figs. \ref{Fig5} (I-III)(b), the temperature there, is higher/less than the temperature of the bottom/top boundary. In other words, while the heat is generated in the hot region, the heat dissipates from the bottom boundary as well. That is why the channel temperature does not increase uncontrollably. In summary, the Figs. \ref{Fig5} (I-III) establish that the top boundary fulfilling the adiabatic condition has always for every material hotter temperature. Also, channels with open boundary conditions are very suitable candidates as the maximum temperature rise can be even for 100 K reduced. 
	
	\section{Conclusions}
	\label{conclusion}
	In this work, two materials of MoSi$_2$N$_4$ and WSi$_2$N$_4$ from the complex MA$_2$Z$_4$ family are thermally investigated. Using the Monte Carlo simulation of the phonon Boltzmann equation, the phonon analysis is performed. Remarkably, the contribution of phonons of each branch on the heat transfer procedure is thoroughly studied. The findings confirmed that WSi$_2$N$_4$, in response to the heat produced in the heating zone, acquires the maximum temperature, almost 100 K less than that of BP and MoSi$_2$N$_4$. The substantial participation of fast and low-frequency TA phonons is introduced as the reason for this very low peak temperature. Also, the presence of fast but high-frequency LA phonons are shown to be the cause of relatively high maximum temperature. As the reliability of the transistors are determined by the peak temperature they reach during the heating, a transistor with a WSi$_2$N$_4$ channel is suggested as the best choice for silicon replacement.
	

\end{document}